\documentclass{PoS}

\usepackage{graphicx}
\usepackage{caption}
\usepackage{subcaption}
\usepackage{subfig}

\title{VERITAS Observations of the Geminga Supernova Remnant}

\ShortTitle{VERITAS Observations of the Geminga Supernova Remnant}

\author{\speaker{Andy Flinders for the VERITAS collaboration}\\
        University of Utah\\
        E-mail: \email{andyflinders@gmail.com}}

\abstract{Geminga was first detected as a gamma-ray point source by the SAS-2 gamma-ray satellite observatory and the COS-B X-ray satellite observatory. Subsequent observations have identified Geminga as a heavily obscured radio-quiet pulsar associated with a nearby (250 pc) late Sedov phase (300,000 year) supernova remnant. The Geminga pulsar is the second brightest source detected by the Large Area Telescope aboard the Fermi gamma-ray satellite (Fermi-LAT) and has been frequently advanced as a source of the anomalous excess of cosmic ray positrons reported by PAMELA, Fermi-LAT, and AMS-2. It is surrounded by a compact X-ray pulsar wind nebula. Observations above 10 TeV by the water Cherenkov observatory Milagro have also revealed a diffuse gamma-ray halo around Geminga extending over several square degrees.  Since 2007 the VERITAS IACT observatory has performed observations of Geminga and the surrounding halo region. However, the standard methods of source detection in VERITAS data have insufficient sensitivity to angularly extended sources (>0.5$^{\circ}$) to reveal a source on the scale of the Milagro detection. In this talk, we describe two approaches being developed to search for angularly extended very high energy gamma-ray emission surrounding the Geminga pulsar.}

\FullConference{The 34th International Cosmic Ray Conference,\\
		30 July- 6 August, 2015\\
		The Hague, The Netherlands}

\begin{document}

\section{Introduction}
Geminga is a nearby (250pc) \cite{2007Ap&SS.308..225F} supernova remnant. Due to its proximity to Earth, it may play a significant role in the nature of our local interstellar environment. The Geminga pulsar has been put forth as the explanation for several astrophysical phenomena including the positron excess measured by PAMELA, Fermi-LAT, and AMS-2 \cite{1996ApJ...461..396M}. The Geminga supernova itself has also been considered as the source of the low density region of the interstellar medium (known as the local bubble \cite{1993Natur.361..706G}) surrounding our Sun. 
Supernova remnants contribute energy to the interstellar medium primarily through the shock front of expanding ejecta and the pulsar wind driven by the spin-down luminosity of the neutron star. Pulsar driven winds commonly exhibit VHE emission e.g. the Crab pulsar wind nebula (PWN) and the Vela PWN. The VERITAS imaging air Cherenkov telescope array has observed the Geminga pulsar location since 2007. Analysis of the VERITAS data using standard point source techniques has yielded no significant detection, and produced a 99\% confidence level limit for a steady/point source at energies above 300 GeV of \begin{math} < 2 * 10^{-12} photons \ cm^{-2} s^{-1} \end{math} \cite{2009arXiv0907.5237F}. A later search for pulsed VHE emission in the VERITAS data also yielded no significant detection \cite{2015ApJ...800...61A}. The Milagro water Cherenkov, observatory which has sensitivity in the range 1 to 100 TeV, has detected very high energy emission from a highly extended (about 4 $^{\circ}$) PWN surrounding the pulsar location \cite{2009ApJ...700L.127A}. The use of established point source analysis techniques is not suitable for analyzing data from potentially highly extended sources. With the development of techniques capable of analyzing highly extended source data it is possible that we may detect extended emission from the Geminga PWN as seen by Milagro. Such techniques will also be useful in analyzing other potentially extended sources.

\section{VERITAS}
The Very Energetic Radiation Imaging Telescope Array System (VERITAS)
is located at the Fred Lawrence Whipple Observatory (FLWO) in southern
Arizona (31$^{\circ}$ 40'N, 110$^{\circ}$ 57'W,  1.3km a.s.l.). The
array consists of four 12-meter Imaging Atmospheric Cherenkov
Telescopes (IACTs) and is sensitive to $\gamma$-rays from
85 GeV to > 35 TeV with an energy resolution of 15-25\%. VERITAS is
able to detect a point source with 1\% the flux of the Crab Nebula
within about 25 hours of observation and has an angular resolution
(68\% containment radius) of better than 0.1$^{\circ}$ at 1 TeV.

When incident on the Earth's atmosphere,
$\gamma$-rays (as well as high energy cosmic rays) induce particle
cascades, or \textit{showers}. The charged component of the shower produces
Cherenkov radiation which is observed by the individual VERITAS
telescope cameras. The resulting camera images are characterized using
Hillas parameters as described in
\cite{1985ICRC....3..445H}. These images are then used to determine
stereoscopic parameters of individual showers, such as the direction
on the sky and core impact position on the ground. Cosmic ray showers
serve as the primary source of background in VERITAS data. Selection cuts
are applied to the combined image parameters to drastically
reduce the number of background cosmic ray events while optimizing 
sensitivity to $\gamma$-ray sources. Additional information
regarding the VERITAS data analysis can be found in 
\cite{2008ICRC....3.1385C,2008ICRC....3.1325D}.

\section{Analysis strategy}
\begin{figure}
\centering
\begin{subfigure}[b]{0.4\textwidth}
                \includegraphics[width=\textwidth]{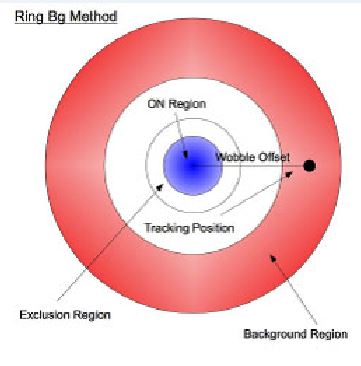}
                \caption{Ring Background Method}
                \label{fig:dSph}
        \end{subfigure}
	\hfill
        ~ 
        \begin{subfigure}[b]{0.4\textwidth}
                \includegraphics[width=\textwidth]{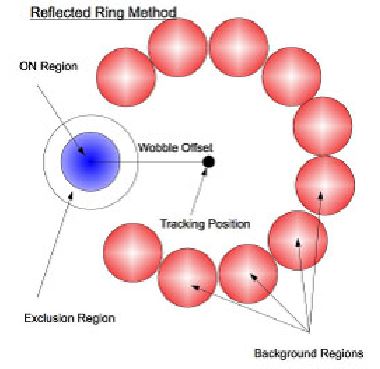}
                \caption{Reflected Region Method}
                \label{fig:Crab}
        \end{subfigure}
\caption{The standard background estimation techniques, the Ring Background Method (left) and the Reflected Region Method (right).}
\label{fig:MatchedTests}
\end{figure}

The ring background model (RBM) and reflected-region model are widely
used to estimate the level of background emission in
IACT data \cite{2007A&A...466.1219B}. This background emission
typically consists of
cosmic ray hadrons and electrons which have passed any selection cuts
applied to the data and are distributed throughout the entire
field of view. The RBM method uses a ring around the
source region to estimate the level of contamination in the region of
interest. The reflected-region requires observations to be taken
offset from the source, then assigns background regions equally offset
from the camera center as the source region. Figure \ref{fig:MatchedTests} gives a visual representation of these techniques. The ratio of the area on the sky designated as the source region to the area designated as the background region is known as $\alpha$. A high value of $\alpha$ means that the contribution of the background estimate's uncertainty to the final uncertainty in the significance of an observation is larger. Because the RBM and RRM require a dedicated background region
within the field of view it is not possible to obtain suitable values of alpha for significantly extended sources using these methods. Furthermore, for significantly extended sources the true
morphology is rarely known, leading to source photons leaking into the
background regions and the obvious drawback of self subtracting the
source under investigation. VERITAS is currently implementing two analysis techniques which have been successfully employed in other VHE experiments to answer the observational challenges presented by extended sources. The first of these techniques is the Matched Run Method (also known as the \textit{ON}/ \textit{OFF} method) which selects a second field of view external to the source and uses it for background estimation \cite{2003ApJ...599..909D}. The second technique is a 3D maximum likelihood method (MLM) similar to the 2D likelihood analysis used in studies of \textit{Fermi}-LAT data \cite{2015ApJS..218...23A}. The MLM presented below includes an added dimension based on a gamma/hadron discriminating parameter to increase sensitivity to extended sources.

\section{Matched Run Method}

Our first approach to extended source analysis is the matched run method. This method attempts to match a given data observation with a second field of view to be used for background estimation. The second field of view needs to have a similar background rate to the \textit{ON} or data observation and is often called the \textit{OFF} observation. This method has been successfully used by previous IACTs such as the Whipple 10m \cite{2003ApJ...599..909D}. This approach allows a source filling the entire field of view of the instrument to be analyzed. To obtain a well matched field of view, ideally an \textit{OFF} observation would be taken immediately before or after the data observation is taken, following the same path in elevation and azimuth. This is undesirable as it requires twice as much observation time compared to the RBM or the RRM. Another way to obtain a good background match is to find a different field of view that has no evidence of $\gamma$-ray emission and has a similar background rate.  

A search algorithm has been developed to help identify which fields of view are suitable background estimation candidates. There are many factors to consider when identifying a possible match for a given \textit{ON} observation, including its elevation angle, azimuthal angle, time of year, hardware configuration, night sky background level, weather, etc. Furthermore, the right ascension and declination of the source can also have a significant effect on the background rate for a source, the primary difference being whether it is on or off of the galactic plane. While all of these factors effect the background rate of IACTs to some degree, several of them are very significant. The elevation angle at which an observation is made has a drastic effect on the number of $\gamma$-rays and cosmic-rays detected, as well as their energy distributions. When searching for a background match for a given data run, elevation angle is the first parameter considered. The second parameter matched in the current algorithm is night sky background. The night sky background level is affected by the right ascension and declination of the observation as well as weather conditions, time of year, and elevation angle and azimuth angle of the observation. Closely matching both elevation angle and night sky background levels between the source and background observations has been effective in finding suitable background regions for several tests thus far. It is important to note that we also require the matched run to have the same hardware configuration of the telescopes, and prefer to use observations recorded as closely together in time as possible. As tests on other sources continue, other restrictions may be required and implemented into the search algorithm. At this time no adjustments are being made to the noise levels of the pixels in order match the observations more closely, as careful matching of the runs has been sufficient.

Once a suitable background observation has been found, there are several ways to analyze the matched pair. One method is to use one data run and one background run directly. The main disadvantage is that only using one background run per \textit{ON} run results in an $\alpha$ of 1, meaning an equal statistical error in the background and signal uncertainty. Using multiple \textit{OFF} observations could decrease the value of $\alpha$ and reduce the statistical uncertainty of the background but comes at the cost of either taking additional \textit{OFF} observations or finding more suitable matched runs. Another way to analyze the pair is to use the \textit{OFF} field of view for a RBM or reflected region analysis. Using a ring (as in the RBM) in the \textit{OFF} field of view can also decrease $\alpha$ and  reduce the statistical uncertainty of the background estimation. Successful development of this technique will be advantageous for several reasons including the ability to analyze previously recorded \textit{ON} data for extended sources, as well as not requiring further \textit{OFF} observations to be taken in the future.

The matched run method is under development and has been used in several test cases.  Testing of the matched run method on fields of view lacking evidence of $\gamma$-ray emission from standard analysis techniques, as well as point source testing, has gone well. Testing on moderately extended sources is underway.

\section{Maximum Likelihood Method}

The maximum likelihood method (MLM) works by modeling the expected
distribution of the data across a set of observables. This includes
modeling both the source and background component of the data. Likelihood analyses
used in other $\gamma$-ray detectors, such as the \textit{Fermi}-LAT
\cite{2015ApJS..218...23A} and ctools for CTA
\cite{2013arXiv1307.5560K}, rely on spatial information alone. For the
\textit{Fermi}-LAT analysis this is sufficient as the instrument also
includes an anticoincidence detector for reducing 
background cosmic ray events by over 99.97\% leading to the
dominant triggers being the result of $\gamma$-ray photons. The
resulting contribution to the \textit{Fermi}-LAT background from
cosmic ray events is small compared to the contribution from
background galactic diffuse photons. For IACT detectors, the
localization of the
photons from point sources is sufficient to implement a purely
spatial MLM. However, for highly extended sources the distribution of
$\gamma$-ray and $\gamma$-ray-like\footnote{\textit{$\gamma$-ray-like} events  typically consist of cosmic ray electrons and/or hadrons
    which survive the standard analysis cuts chosen
    to optimize sensitivity to $\gamma$-ray sources. These events are
    the main source of background in VERITAS data.} events can be
  quite similar leading to a reduced sensitivity of the analysis. To
  overcome this, the MLM in development for VERITAS data analysis
  includes a third dimension based on a $\gamma$/hadron discriminating
  parameter known as mean scaled width (MSW). This third dimension is
  incorporated into the likelihood by including models of how both the
  $\gamma$-ray and background cosmic ray components of our data are
  distributed in MSW. Details regarding
the VERITAS MLM, including information on the likelihood construction,
background spatial model determination, and characterization of the
various IRFs can be found in \cite{MLM_ICRC2015Proc}.

\begin{figure}
\centering
        \begin{subfigure}[b]{0.49\textwidth}
\centering
                \includegraphics[width=\textwidth]{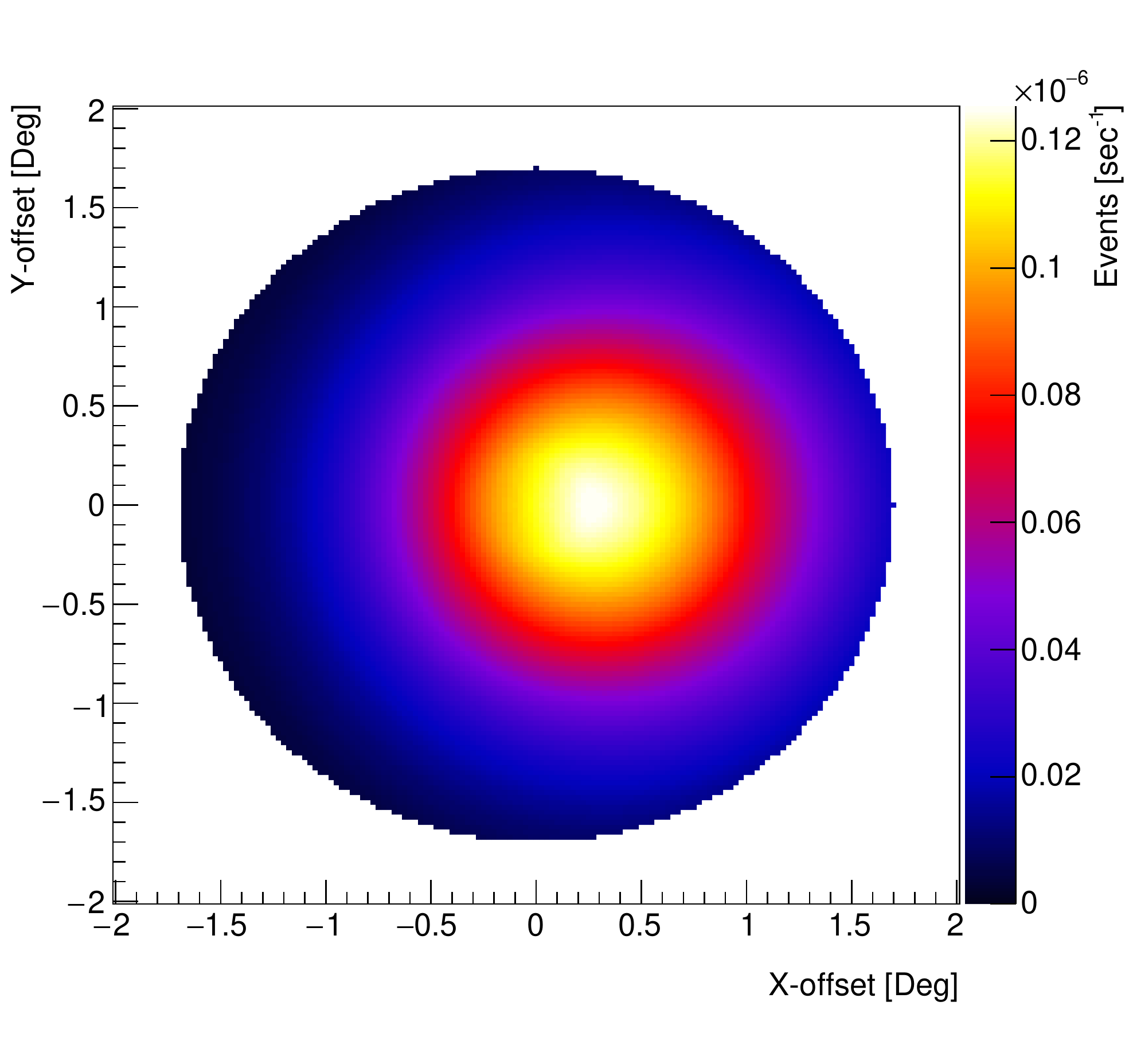}
                \caption{Extended source spatial model}
                \label{fig:ExtendedSourceModel}
        \end{subfigure}
        ~ 
        \begin{subfigure}[b]{0.49\textwidth}
                \includegraphics[width=\textwidth]{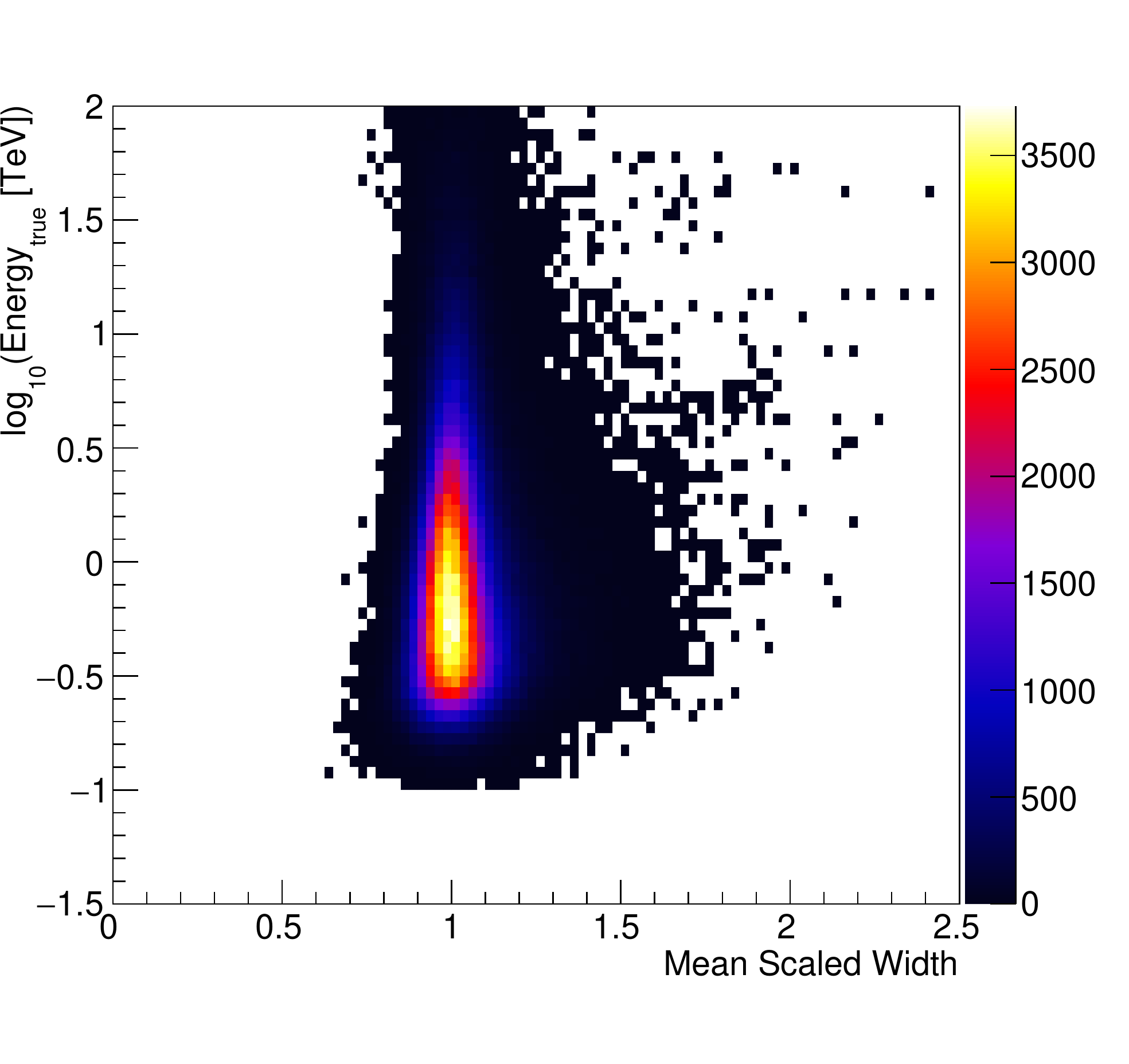}
                \caption{$\gamma$-ray MSW distribution}
                \label{fig:BackgroundMSW}
        \end{subfigure}

        \begin{subfigure}[b]{0.49\textwidth}
                \includegraphics[width=\textwidth]{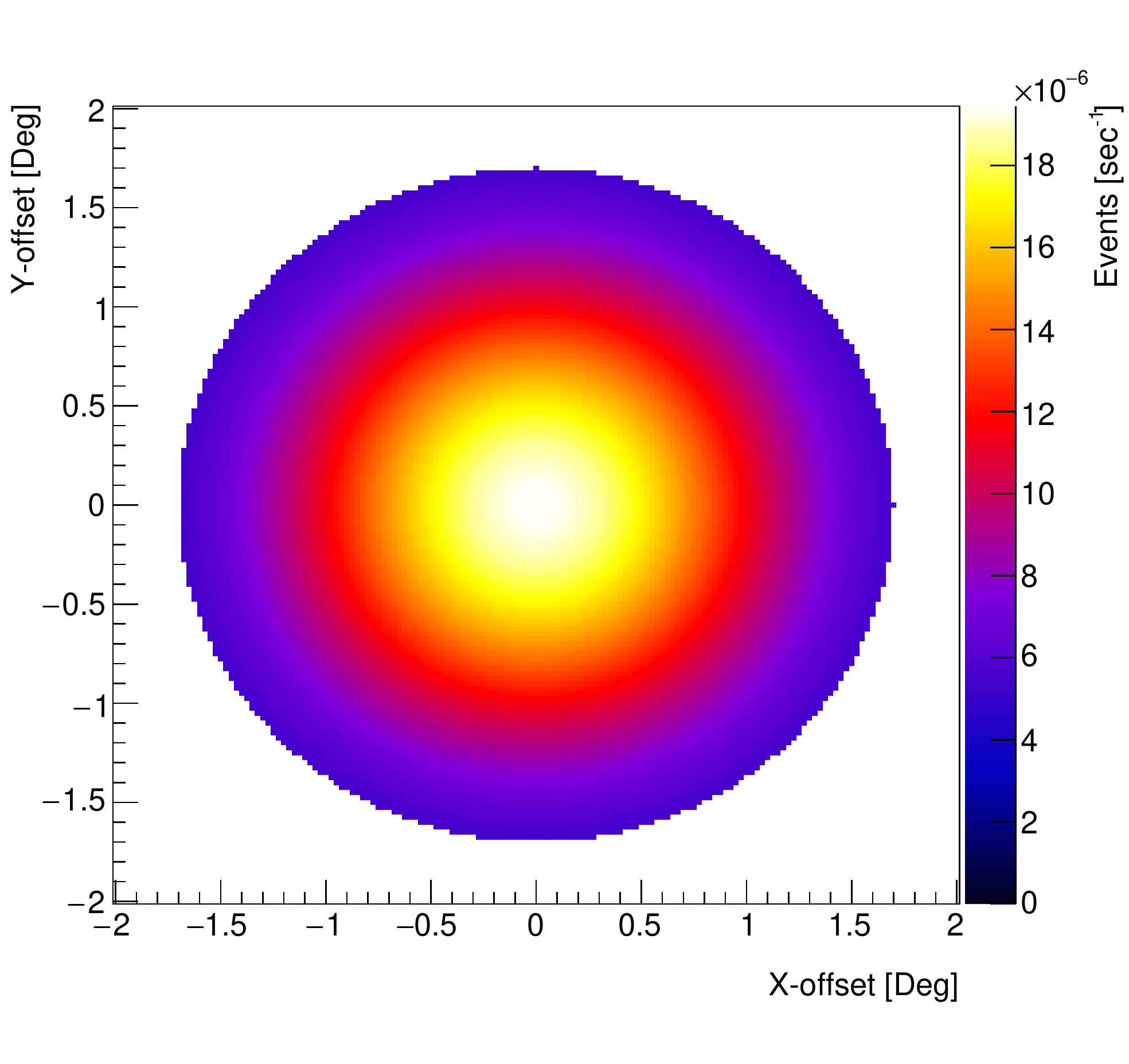}
                \caption{Background spatial model}
                \label{fig:SourceMSW}
        \end{subfigure}
        ~ 
        \begin{subfigure}[b]{0.49\textwidth}
                \includegraphics[width=\textwidth]{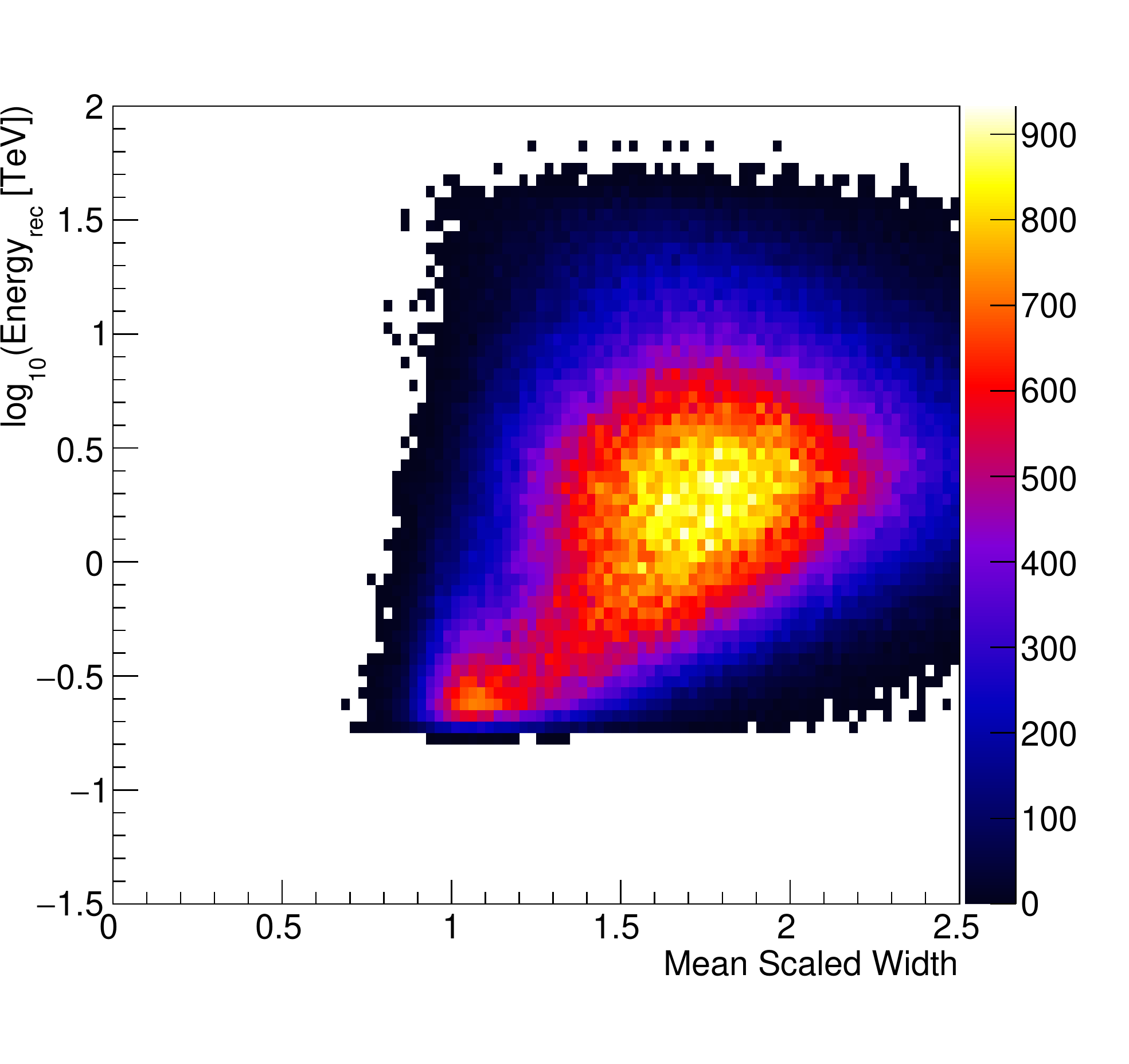}
                \caption{Background MSW distribution}
                \label{fig:BackgroundMSW}
        \end{subfigure}
\caption{(a) Extended source spatial model as derived from the Milagro
  C3 source parameters 
  described in \cite{2009ApJ...700L.127A}. This model is asymmetric
  due to the observation offset of 1$^{\circ}$ and the changing
  sensitivity of the instrument over the field of view. (b) MSW
  distribution derived from $\gamma$-ray air-shower simulations.(c)
  Background spatial model and (d) background MSW model derived from
  VERITAS data. Both background distributions are derived from events
  greater than 0.4$^{\circ}$ from known and potential $\gamma$-ray
  sources. Additionally, the spatial background distribution uses
  events with MSW < 1.3 reflecting the range currently used in the MLM
  analysis. Distributions were derived assuming 70$^{\circ}$
  elevation, due south, with four telescopes participating in the
  shower reconstruction.}
\label{fig:MLMDistributions}
\end{figure}

MSW is determined as an average of the Hillas width parameters from each
telescope image normalized by the expected width derived from
simulated air-showers with the same integrated image intensity and shower core impact
distance from the telescope.
\begin{equation}\label{eq:MSW}
MSW=\left(\frac{1}{n}\right)\sum_{i=1}^{n}\frac{w_{i}}{<w_{i}>}
\end{equation}
It can be seen from Eq. \ref{eq:MSW} that the MSW distribution for
$\gamma$-ray events is expected to peak at $\sim$1. Cosmic ray events
typically produce broader showers\footnote{This is due to the larger
  transverse momentum of the secondary pions produced in hadronic cosmic ray showers.}
resulting in images with much larger
calculated widths. This leads to a peak at higher values of MSW
\cite{1985ICRC....3..445H} as compared to $\gamma$-ray showers (see
Figure \ref{fig:MLMDistributions}). Of the standard set of selection
cuts used to reduce the level of background in VERITAS data, a cut on
MSW is among the most efficient. For a typical VERITAS source analysis, events
with MSW>1.1 are cut. However, in the MLM analysis the MSW
range is extended to 1.3 to provide ample separation between the
$\gamma$-ray and cosmic ray MSW distributions. The distinct difference
in the shape of the MSW distribution for these two components improves
the MLM's ability to distinguish source emission from background as
opposed to using a spatial only model.

The source models are based on the instrument response functions
(IRFs) relevant to the VERITAS detector. The IRFs are combined based on the
formulation of Mattox (1996) Eq. 2
\cite{1996ApJ...461..396M}. The actual construction of the source model
is made more tractable by computing the expected spatial models in
0.025$^{\circ}$$\times$0.025$^{\circ}$ bins. Each IRF component is
also computed in narrow logarithmic bins of true
energy\footnote{\textit{true energy} is the true energy of an observed
$\gamma$-ray. Observed (or reconstructed) energy is the energy of the
photon as calculated by the VERITAS analysis.} in steps of
0.05. The final source model for a given spatial bin is then computed
according to
\begin{equation}
\label{eq:MLMSourceModel}
S_{src}(\mathbf{\vec{r}}_{i,j}|\mathbf{\vec{s}})=\sum_{k,m,n}B_{m,n}P_{m,n}(\mathbf{\vec{r}}_{i,j})A_{m,n}(E_{k}')\int_{E_{min}}^{E_{max}}R_{m,n}(E,E_{k}')dE\int_{E'_{k,lower}}^{E'_{k,upper}}S(E'|\mathbf{\vec{s}})dE'
\end{equation}
where the components consist of the intrinsic source
morphology (\textit{B}), the instrument point spread function
(\textit{P}), effective collection area (\textit{A}), energy
dispersion (\textit{R}), and source spectrum (\textit{S}). Spatial
coordinates are given by $\mathbf{\vec{r}}_{i,j}$, \textit{E'}
represents the true energy of an event, and \textit{E} the observed
energy. To account for the smearing of events due to our point spread function, the
contriubtion from every bin in the map (indexed by
\textit{m},\textit{n}) is summed at the position of interest (indexed
by \textit{i},\textit{j}). Each IRF is parameterized for a given set of observing
conditions in zenith angle, azimuth, source offset, and camera noise
level. The IRFs are also computed in narrow bins of \textit{E'},
indexed by \textit{k}. Integrating the computed source model over
every bin (\textit{i,j}) yields an estimate of the total number of
observed photons in a given dataset. 

To demonstrate how the inclusion of the MSW parameter aids in the
detection of extended sources, a comparison of the models of an
extended source were compared to those of the expected background (see
Figure \ref{fig:MLMDistributions}). The
extended source uses as input the extension and spectrum as reported in
\cite{2009ApJ...700L.127A} for the Milagro C3 source. The morphology
(\textit{B}) is assumed to be a Gaussian with $\sigma$=1.3$^{\circ}$ and the
spectrum is modeled as a powerlaw with spectral index of -2.6 and flux
of 37.7$\times$10$^{-17}$ (TeV$^{-1}$cm$^{-2}$s$^{-1}$) at 35
TeV. The resulting source model was then derived for the VERITAS
energy range 0.5 - 1 TeV. When compared to the expected background
spatial model, the derived extended source model is very
similar. However, in the MSW dimension there is a significant
separation between the two models.

The presented extrapolation of the Milagro C3
source to IACT energies should only be considered a test of the
MLM. In reality, due to the higher energy threshold and larger
uncertainty in the event position reconstruction of the Milagro
detector, any emission potentially seen at VERITAS energies may have a
very different morphology and spectrum.

\section{Conclusions} 

The Geminga pulsar is an interesting $\gamma$-ray source with a rich discussion surrounding it in connection with our stellar neighborhood. A VERITAS detection of either a point source or an extended PWN would contribute significantly to our understanding of the origin and evolution of the local interstellar medium. Two analysis techniques are being implemented by VERITAS which will aid in the study of spatially extended $\gamma$-ray emission which may emanate from this region. These techniques are both progressing well and should be ready to analyze the VERITAS Geminga data in the near future. 

\section{Acknowledgments}

This research is supported by grants from the U.S. Department of Energy Office of Science, the U.S. National Science Foundation and the Smithsonian Institution, and by NSERC in Canada. We acknowledge the excellent work of the technical support staff at the Fred Lawrence Whipple Observatory and at the collaborating institutions in the construction and operation of the instrument. The VERITAS Collaboration is grateful to Trevor Weekes for his seminal contributions and leadership in the field of VHE gamma-ray astrophysics, which made this study possible. 


\bibliographystyle{JHEP}
\bibliography{Geminga_MLMComponent}{}

\providecommand{\href}[2]{#2}\begingroup\raggedright\begin{thebibliography}{10}

\bibitem{2007Ap&SS.308..225F}
J.~{Faherty}, F.~M. {Walter}, and J.~{Anderson}, {\it {The trigonometric
  parallax of the neutron star Geminga}},  {\em Astrophysics and Space Science}
  {\bf 308} (Apr., 2007) 225--230.

\bibitem{1996ApJ...461..396M}
J.~R. {Mattox} et~al., {\it {The Likelihood Analysis of EGRET Data}},  {\em
  ApJ} {\bf 461} (Apr., 1996) 396.

\bibitem{1993Natur.361..706G}
N.~{Gehrels} and W.~{Chen}, {\it {The Geminga supernova as a possible cause of
  the Local Interstellar Bubble}},  {\em Nature} {\bf 361} (Feb., 1993) 706.

\bibitem{2009arXiv0907.5237F}
G.~{Finnegan} and {for the VERITAS Collaboration}, {\it {Search for TeV
  Emission from Geminga by VERITAS}},  {\em ArXiv e-prints} (July, 2009)
  [\href{http://arxiv.org/abs/0907.5237}{{\tt arXiv:0907.5237}}].

\bibitem{2015ApJ...800...61A}
E.~{Aliu} et~al., {\it {A Search for Pulsations from Geminga above 100 GeV with
  VERITAS}},  {\em The Astrophysical Journal} {\bf 800} (Feb., 2015) 61,
  [\href{http://arxiv.org/abs/1412.4734}{{\tt arXiv:1412.4734}}].

\bibitem{2009ApJ...700L.127A}
A.~A. {Abdo} et~al., {\it {Milagro Observations of Multi-TeV Emission from
  Galactic Sources in the Fermi Bright Source List}},  {\em The Astrophysical
  Journal} {\bf 700} (Aug., 2009) L127--L131,
  [\href{http://arxiv.org/abs/0904.1018}{{\tt arXiv:0904.1018}}].

\bibitem{1985ICRC....3..445H}
A.~M. {Hillas}, {\it {Cerenkov light images of EAS produced by primary gamma}},
   {\em Proceedings of the 19th International Cosmic Ray Conference} (Aug.,
  1985) 445--448.

\bibitem{2008ICRC....3.1385C}
P.~{Cogan}, {\it {VEGAS, the VERITAS Gamma-ray Analysis Suite}},  {\em
  Proceedings of the 30th International Cosmic Ray Conference} {\bf 3} (2008)
  1385--1388, [\href{http://arxiv.org/abs/0709.4233}{{\tt arXiv:0709.4233}}].

\bibitem{2008ICRC....3.1325D}
M.~K. {Daniel}, {\it {The VERITAS standard data analysis}},  {\em Proceedings
  of the 30th International Cosmic Ray Conference} {\bf 3} (2008) 1325--1328,
  [\href{http://arxiv.org/abs/0709.4006}{{\tt arXiv:0709.4006}}].

\bibitem{2007A&A...466.1219B}
D.~{Berge}, S.~{Funk}, and J.~{Hinton}, {\it {Background modelling in
  very-high-energy {$\gamma$}-ray astronomy}},  {\em A\&A} {\bf 466} (May,
  2007) 1219--1229, [\href{http://arxiv.org/abs/astro-ph/0610959}{{\tt
  astro-ph/0610959}}].

\bibitem{2003ApJ...599..909D}
I.~{de la Calle P{\'e}rez} et~al., {\it {Search for High-Energy Gamma Rays from
  an X-Ray-selected Blazar Sample}},  {\em The Astrophysical Journal} {\bf 599}
  (Dec., 2003) 909--917, [\href{http://arxiv.org/abs/astro-ph/0309063}{{\tt
  astro-ph/0309063}}].

\bibitem{2015ApJS..218...23A}
F.~{Acero} et~al., {\it {Fermi Large Area Telescope Third Source Catalog}},
  {\em ApJS} {\bf 218} (June, 2015) 23,
  [\href{http://arxiv.org/abs/1501.2003}{{\tt arXiv:1501.2003}}].

\bibitem{2013arXiv1307.5560K}
J.~{Kn{\"o}dlseder} et~al., {\it {Towards a common analysis framework for
  gamma-ray astronomy}},  {\em Proceedings of the 33rd International Cosmic Ray
  Conference} (July, 2013) [\href{http://arxiv.org/abs/1307.5560}{{\tt
  arXiv:1307.5560}}].

\bibitem{MLM_ICRC2015Proc}
J.~V. {Cardenzana for the VERITAS Collaboration}, {\it {A Novel Method for
  Detecting Extended Sources with VERITAS}},  {\em Proceedings of the 34th
  International Cosmic Ray Conference} (2015).

\end{thebibliography}\endgroup

\end{document}